\documentclass[twocolumn]{aastex631}

\newcommand{\angstrom}{\textup{\AA}}

\newcolumntype{b}{X}
\newcolumntype{s}{>{\hsize=.5\hsize}X}

\usepackage{textcomp}
\usepackage{CJKutf8}
\usepackage{amsmath}
\usepackage{color,soul}
\usepackage{tabularx}

\shorttitle{Resolution-Corrected White Dwarf Gravitational Redshifts}
\shortauthors{Arseneau et al.}

\begin{document}

\title{Resolution-Corrected White Dwarf Gravitational Redshifts Validate SDSS-V Wavelength Calibration and Enable Accurate Mass-Radius Tests}%For Accurate Tests of Mass-Radius Relationships}

\correspondingauthor{Stefan M. Arseneau}
\author[0000-0002-6270-8624]{Stefan M. Arseneau}
\email{arseneau@bu.edu}
\affiliation{Department of Astronomy \& Institute for Astrophysical Research, Boston University, 725 Commonwealth Ave., Boston, MA 02215, USA}

\author[0000-0001-5941-2286]{J.~J.~Hermes}
\affiliation{Department of Astronomy \& Institute for Astrophysical Research, Boston University, 725 Commonwealth Ave., Boston, MA 02215, USA}

\author[0000-0001-6100-6869]{Nadia L. Zakamska}
\affiliation{William H. Miller III Department of Physics and Astronomy, Johns Hopkins University, Baltimore, MD 21210, USA}

\author[0000-0002-6871-1752]{Kareem El-Badry}
\affiliation{Department of Astronomy, California Institute of Technology, 1200 East California Boulevard, Pasadena, CA 91125, USA}

\author[0000-0002-8866-4797]{Nicole R. Crumpler}
\altaffiliation{NSF Graduate Research Fellow}
\affiliation{William H. Miller III Department of Physics and Astronomy, Johns Hopkins University, Baltimore, MD 21210, USA}

\author[0000-0002-0572-8012]{Vedant Chandra}
\affiliation{Center for Astrophysics $\mid$ Harvard \& Smithsonian, 60 Garden St, Cambridge, MA 02138, USA}

\author[0000-0002-5864-1332]{Gautham Adamane Pallathadka}
\affiliation{William H. Miller III Department of Physics and Astronomy, Johns Hopkins University, Baltimore, MD 21210, USA}

\author[0000-0003-3494-343X]{Carles Badenes}
\affiliation{Department of Physics and Astronomy, University of Pittsburgh, 3941 O’Hara Street, Pittsburgh, PA 15260, USA}
\affiliation{Pittsburgh Particle Physics, Astrophysics, and Cosmology Center (PITT PACC), University of Pittsburgh, Pittsburgh, PA 15260, USA}

\author[0000-0002-2761-3005]{Boris T. G\"{a}nsicke}
\affiliation{Astronomy and Astrophysics Group, Department of Physics, University of Warwick, Coventry, CV4 7AL, United Kingdom}

\author[0000-0002-6428-4378]{Nicola Gentile Fusillo}
\affiliation{Department of Physics, Universita' degli Studi di Trieste, Via A. Valerio 2, 34127, Trieste, Italy}
\affiliation{INAF-Osservatorio Astronomico di Trieste, Via G.B. Tiepolo 11, I-34143 Trieste, Italy}
\begin{abstract}

\noindent

Leveraging the large sample size of low-resolution spectroscopic surveys to constrain white dwarf stellar structure requires an accurate understanding of the shapes of hydrogen absorption lines, which are  pressure broadened by the Stark effect. Using data from both the Sloan Digital Sky Survey and the Type Ia Supernova Progenitor Survey, we show that substantial biases ($5-15$~km s$^{-1}$) exist in radial velocity measurements made from observations at low spectral resolution relative to similar measurements from high-resolution spectra. Our results indicate that the physics of line formation in high-density plasmas, especially in the wings of the lines, are not fully accounted for in state-of-the-art white dwarf model atmospheres. We provide corrections to account for these resolution-induced redshifts in a way that is independent of an assumed mass-radius relation, and we demonstrate that statistical measurements of gravitational redshift with these corrections yield improved agreement with theoretical mass-radius relations. Our results provide a set of best practices for white dwarf radial velocity measurements from low-resolution spectroscopy, including those from the Sloan Digital Sky Survey, the Dark Energy Spectroscopic Instrument, the 4-meter Multi-Object Spectroscopic Telescope, and the Wide-Field Multiplexed Spectroscopic Facility.

\end{abstract}

\section{Introduction} \label{sec:intro}

White dwarfs are the endpoint of stellar evolution for the vast majority of stars. Having exhausted the fuel necessary to sustain fusion, these retired stars are supported by electron degeneracy pressure. White dwarfs have a sharply peaked mass distribution around a mean mass of $\approx 0.6$~$M_\odot$, corresponding to a typical radius of $\approx 1.40$~$R_\oplus$ \citep{2007MNRAS.375.1315K}. With no means to actively generate heat, they spend the rest of their lives cooling monotonically from temperatures $>150,000$~K to $<4000$~K \citep{2001PASP..113..409F, 2010A&ARv..18..471A}. The majority of white dwarfs have spectra dominated by the presence of hydrogen absorption lines \citep{2020ApJ...901...93B} which are strongly pressure-broadened due to the first-order expansion of the Stark effect. 

Because of their high temperatures and compact nature, white dwarfs are promising observational laboratories for material under conditions which are otherwise rare or difficult to reproduce experimentally \citep{2022PhR...988....1S}. Their compactness gives rise to a substantial gravitational redshift: a shift caused by the energy needed for a photon to leave the star's potential well. The average white dwarf gravitational redshift is $32$~km s$^{-1}$ for a $0.6~M_\odot$ white dwarf \citep{2010ApJ...712..585F}. Gravitational redshift is a sensitive function of stellar mass, ranging from $\approx15$~km s$^{-1}$ at $0.4~M_\odot$ to $>200$~km s$^{-1}$ for the most massive white dwarfs \citep{2020ApJ...901...93B}.  

It is difficult to measure white dwarf gravitational redshifts because a white dwarf's apparent radial velocity, that which is measured from spectroscopy, consists of both its gravitational redshift and redshift due to its true radial motion through space. When this degeneracy can be broken, gravitational redshifts are a powerful probe of white dwarf stellar structure. \cite{2019A&A...627L...8P} subtracted out the mean motion of the Hyades cluster to measure the gravitational redshifts of six white dwarfs, and \cite{2024ApJ...963...17A} and \cite{2025A&A...695A.131R} used samples of 135 and 50 white dwarfs with main-sequence binary companions respectively to measure individual white dwarfs' gravitational redshifts. \cite{2020ApJ...899..146C} and \cite{2024ApJ...977..237C} used a large sample of radial velocity measurements from the Sloan Digital Sky Survey (SDSS) to statistically measure gravitational redshifts from thousands of isolated white dwarfs. These studies have provided strong validation of theoretical calculations of the white dwarf mass-radius relation.

Some of the first measurements of gravitational redshifts in white dwarfs were made by \cite{1967ApJ...149..283G} and \cite{1972ApJ...177..441T}, who noted radial velocities substantially higher than those predicted by theoretical mass-radius relations: systematically redshifted by $10-15$~km s$^{-1}$. In response, \cite{PhysRevA.6.1132} carried out laboratory experiments measuring hydrogen Balmer emission lines. They found additional systematic redshifted asymmetries caused by nonlinear (specifically quadratic) terms in the expansion of the Stark effect in dense plasmas. This was soon validated by \cite{1975A&A....45..159G} and \cite{1987ApJ...313..750G}, who found that the magnitude of the pressure shift was proportional to the ratio of electron density to temperature. Contemporary experiments using the Z-pinch pulsed power facility at Sandia National Laboratory are attempting to extend the work of \cite{PhysRevA.6.1132} to a parameter space more representative of the physical conditions of white dwarfs \citep{2016PhRvA..94b2501G, 2020PhPl...27g0501S, 2022ApJ...927...70C}.

%\subsection{Hydrogen Balmer Line Shapes}

Hydrogen absorption lines found in white dwarf spectra consist of two parts: a central Gaussian-shaped core within $\approx 3$~$\angstrom$ of the line centroid caused by non-local thermodynamic equilibrium (NLTE) Doppler broadening due to thermal motion, and a wide local thermodynamic equilibrium (LTE) Lorentzian-shaped wing extending up to $100~\angstrom$ from the line centroid created by pressure broadening. The wings are symmetrical in energy space to first order, and therefore exhibit a small asymmetry in wavelength space \citep{1988ApJ...331..794H, 2009ApJ...696.1755T, 2010MmSAI..81..921K}.

As the location of maximum hydrogen opacity, the NLTE core forms high in the stellar photosphere where pressures and temperatures are the lowest. Radial velocity measurements using the NLTE core consist only of those effects induced by the white dwarf's random actual space motion ($v_\text{Space}$) and its gravitational redshift ($v_\text{g}$):
\[v_\text{Core} = v_\text{Space} + v_\text{g}\]
and are accurate tracers of the system's true motion \citep{2020A&A...638A.131N}. 

The LTE wings form in deeper regions of the star where higher-order effects of physical processes such as pressure shifts or Coulomb interactions are more important. This means that the apparent radial velocity measured from low-resolution spectroscopy which cannot resolve the core could contain additional unmodeled signals arising from the treatment of these terms in line shape models:
\[v_\text{Wing} = v_\text{Space} + v_\text{g} + v_\text{Resolution}.\]
By comparing the radial velocity measured from low-resolution spectra which are dominated by the wings to that measured from spectra which resolve the core, resolution-induced redshift terms arising from the wings can be isolated.

\cite{2025A&A...695A.131R} searched for resolution-induced redshift terms by downgrading the spectral resolution of their observations to 2\,\AA\ and fitting radial velocities the first five Balmer lines. They found a systematic shift of $v=2.9\pm 11.8$~km s$^{-1}$ compared to that measured from high-resolution data.%, translating to a systematic uncertainty in the white dwarf mass distribution of $M=0.06\pm0.23$~$M_\odot$.}

The photospheres of hot, hydrogen-rich (spectral type DA) white dwarfs consist almost exclusively of dense and fully ionized hydrogen plasma. As white dwarfs cool to $\approx 18,000~K$ (the onset of partial hydrogen ionization; \citealt{2019MNRAS.488.2503C}), DA photospheres develop a surface convective zone. Convection is initially inefficient until the star cools to $< 12,000~K$ \citep{2015ASPC..493...89T}. Hydrogen recombination is also associated with the blue edge of the ZZ Ceti instability strip \citep{2015A&A...575A.125V}, and with changes in the temperature and electron density profiles of the photosphere \citep{2022ApJ...927..128B,2022PhR...988....1S}. These effects are computationally difficult to model and have the potential to introduce unmodeled sources of bias in fits to spectral lines \citep{2015ApJ...799..142T}.

Modern calculations of line shapes in white dwarfs are based on the theoretical work of \cite{1970JQSRT..10.1011V}. Despite the remarkable success of their work and the line profiles derived from it in reproducing observational data \citep{1988ApJ...331..794H, 2009ApJ...696.1755T}, their work only treats the Stark effect to first order and does not take into account asymmetries due to higher-order perturbative effects or non-dipolar electrodynamic interactions between charged particles in the atmosphere. The increase in quantity of white dwarf spectroscopic data associated with surveys such as the fifth generation of the Sloan Digital Sky Survey (SDSS-V; \citealt{2025arXiv250706989K}) and the Dark Energy Spectroscopic Instrument (DESI; \citealt{2024AJ....168...58D}) have emphasized the need for the consideration of higher-order effects in white dwarf model atmospheres.

In this work, we use high-resolution spectroscopic data from the Type Ia Supernova Progenitor Survey (SPY; \citealt{2020A&A...638A.131N}) and low-resolution spectroscopic data from SDSS to measure a resolution-induced, unmodeled radial velocity signal as a function of effective temperature. In Section \ref{sec:observations} we describe the SPY and SDSS datasets. In Section \ref{sec:high-res} we describe our methods for using the high-resolution SPY dataset to perform a direct measurement of resolution-induced redshifts independent of assumptions about mass-radius relations. In Section \ref{sec:low-res} we use the low-resolution SDSS dataset and an assumed mass-radius relation to measure this effect statistically with improved effective temperature fidelity. We present our results in Section \ref{sec:results} and discuss them in Section \ref{sec:discuss}. Throughout the paper, we adopt median statistics to minimize the effects of outliers.

\section{Data and Observations} \label{sec:observations}

\subsection{High-Resolution Spectroscopic Data from SPY}

The SPY survey operated from 2002 to 2020 using the Ultraviolet and Visual Echelle Spectrograph (UVES) on the European Southern Observatory's $8.2$~meter Very Large Telescope (ESO VLT; \citealt{2000SPIE.4008..534D}). SPY spectra are taken in individual 30 minute exposures with a slit width of $2.1$~arcseconds in order to maximize incident flux on the detector for faint targets. With this configuration, SPY attains a spectral resolution of $R=18,500$ (or 0.35\,\angstrom\ at H$\alpha$) which is sufficient to resolve the core of white dwarf Balmer lines from which we can measure the most unbiased radial velocities.

\cite{2020A&A...638A.131N} provides 1391 spectra of 643 DA white dwarfs as well as their measured apparent radial velocities. After removing known binaries and magnetic white dwarfs using Table C2 of \cite{2020A&A...638A.131N}, we are left with 1248 individual exposures of 578 DA white dwarfs. These spectra are provided in air wavelengths and in the rest frame of the observer. In order to attain maximum agreement with the radial velocities measured by \cite{2020A&A...638A.131N} we first use \texttt{astropy} \citep{astropy:2013, astropy:2018, astropy:2022} to apply a Doppler shift to each spectrum, bringing it into the heliocentric reference frame using the location and Julian Date of each exposure. Next, we transform the reported wavelengths from air to vacuum \citep{1991ApJS...77..119M}. These steps in this order ensure that different exposures of the same object are comparable and maximize agreement with the radial velocities reported by \cite{2020A&A...638A.131N}.

The spectra associated with \cite{2020A&A...638A.131N} do not include estimates of uncertainties for given flux measurements. We estimate the signal to noise ratio per pixel of each observation by measuring the ratio of mean measurement to standard deviation in the wavelength range of $5260~\angstrom < \lambda < 5280~\angstrom$. This region is devoid of emission or absorption features in DA white dwarfs. We then take the uncertainty of each flux measurement to be that associated with our estimated signal-to-noise ratio.

Finally, we cross-match the SPY sample against astrometry from \textit{Gaia} \citep{2016A&A...595A...1G, 2023A&A...674A...1G} and stellar parameters of $T_\text{eff}$ and $\log g$ measured assuming a hydrogen-dominated atmosphere by \cite{2021MNRAS.508.3877G}. We choose to use photometric stellar parameters rather than spectroscopic values following \cite{2024ApJ...969...68H}, who found that photometric parameters yielded greater internal consistency in the total ages of wide ($>$100\,au) WD+WD binary systems. Photometric parameters are also less sensitive to input physics in the calculation of spectral line shapes. We restrict our sample to those objects with \texttt{parallax\_over\_error > 3} and \texttt{ruwe < 1.20}, the latter recommended by \cite{2025AJ....169...29S} to reliably filter out marginally resolved binaries, leaving us with 1170 exposures of 543 DA white dwarfs. 

\subsection{SDSS Spectroscopic Data}

Our statistical analysis is based on the Sloan Digital Sky Survey's DR19 Value Added Catalog (VAC) of $26,041$ spectroscopically identified DA white dwarfs and measured parameters (VAC; \citealt{vac_paper}). The ongoing fifth generation of SDSS (SDSS-V; \citealt{2025arXiv250706989K}) started operations in November 2020 using the Apache Point Observatory 2.5m telescope \citep{2006AJ....131.2332G}. Targets in the VAC were observed with the Baryon Oscillation Spectroscopic Survey spectrograph (BOSS; \citealt{2013AJ....146...32S}) using reduction pipeline v6\_1\_3. BOSS covers a wavelength range of $3650-9500$~$\angstrom$ at a spectral resolution of $R\approx1800$. The VAC is publicly available, having been released as part of SDSS DR19\footnote{\url{https://www.sdss.org/dr19/data_access/value-added-catalogs/?vac_id=10008}} \citep{2025arXiv250707093S}. 

The VAC contains cross-matches of $T_{\text{eff}}$ and $\log g$ measurements made by \cite{2021MNRAS.508.3877G}. Additionally, it contains $T_\text{eff}$, $\log g$, and radial velocity measurements produced from spectroscopic data via the python package \texttt{corv} \footnote{\url{https://github.com/vedantchandra/corv}} \citep{2024ApJ...963...17A} with 3D model atmospheres \citep{2009ApJ...696.1755T, 2013A&A...559A.104T, 2015ApJ...809..148T}. For our analysis, we adopt the $T_\text{eff}$ and $\log g$ values of \cite{2021MNRAS.508.3877G}. To ensure that our dataset contains accurate radial velocities, we select only those white dwarfs that have co-added SDSS spectra with signal-to-noise ratio greater than 10, and we adopt the same \textit{Gaia} \texttt{parallax\_over\_error} and \texttt{ruwe} cuts as with the SPY sample. Additionally, the VAC contains a parameter $\eta$ representing the likelihood of an object with multiple epochs of data being a binary based on radial velocity variability, originally described by \cite{2000MNRAS.319..305M} and \cite{2017MNRAS.468.2910B}. Following their recommendation, we ignore objects with $\eta>2.5$ to remove possible binaries.

Additionally, we require that the spectroscopic $\log g$ reported by \texttt{corv} be no less than 7.01 and no greater than 8.99. Finally, we require that the mass measured from the spectral energy distribution (SED) be greater than $0.35$~$M_\odot$ (via the VAC column \texttt{mass\_rad\_theory}). The latter filters out extremely low mass (ELM) white dwarfs, which are required by the current understanding of stellar evolution to form in binaries \citep{2020ApJ...889...49B}. These steps will not filter out all white dwarf binaries, but they ensure that the remaining systems have minimal radial-velocity variability.

\subsubsection{Radial Velocity Offsets in Legacy BOSS Data}

%\begin{figure}
%    \centering
%    \includegraphics[width=\linewidth]{figures/offset_teff.pdf}
%    \caption{Difference between radial velocities measured from SDSS-V and those from earlier versions of the survey (eSDSS) plotted against effective temperatures measured from eSDSS photometry. The best fit line is consistent with no temperature dependence, with slope $(0.7 \pm 1.0)\times 10^{-4}$~km s$^{-1}$ K$^{-1}$ and intercept $6.0\pm 1.6$~km s$^{-1}$, implying a systematic blueshift of eSDSS spectra. The weighted mean offset is $7.1\pm0.7$~km s$^{-1}$.}
%    \label{fig:offset}
%\end{figure}

As noted by \cite{vac_paper}, the wavelength calibration of legacy BOSS spectra is inconsistent with that of SDSS-V. We investigate the magnitude of the difference in a subset of $561$ stars in our sample which have radial velocities measured from co-added spectra in both SDSS-V and previous generations of the survey using the same spectrograph. We do not find a significant temperature dependence, with a best fit slope of $(0.7 \pm 1.0)\times 10^{-4}$~km s$^{-1}$ K$^{-1}$ and intercept of $6.0\pm 1.6$~km s$^{-1}$.

The detector associated with the BOSS instrument is split into two channels, with H$\alpha$ (the hydrogen absorption line with the strongest NLTE core) falling on the red channel and the other lines falling on the blue channel \citep{2013AJ....145...10D}. We further investigate the consistency in wavelength calibration by fitting radial velocities to the H$\alpha$, H$\beta$, H$\gamma$, and H$\delta$ lines separately for both the SDSS-V and legacy BOSS spectra of all $561$ stars using the 3D DA LTE model spectra\footnote{\url{https://warwick.ac.uk/fac/sci/physics/research/astro/people/tremblay/modelgrids}} of \cite{2013A&A...559A.104T, 2015ApJ...809..148T}. We find the median offset between SDSS-V and legacy spectra is $-4.8\pm0.4$~km s$^{-1}$ using H$\alpha$, $7.1\pm0.2$~km s$^{-1}$ using H$\beta$, $15.3\pm0.3$~km s$^{-1}$ using H$\gamma$, and $22.8\pm0.3$~km s$^{-1}$ using H$\delta$. Because the deviation between the two radial velocity distributions becomes greater with bluer lines, we conclude that it is due to improvements in wavelength calibrations for the blue channel rectified in SDSS-V. 

We correct all VAC radial velocities measured from legacy BOSS spectra. VAC radial velocities are measured from simultaneous fits of the first four Balmer lines. Because we do not detect any temperature dependence, we add to them the mean offset weighted by uncertainties, $7.1\pm0.7$~km s$^{-1}$. We demonstrate in Section \ref{sec:practices} why we believe that the SDSS-V wavelength solution is superior.

\section{Direct Measurements of Resolution-Induced Redshifts} \label{sec:high-res}

\subsection{High-Resolution Apparent Radial Velocities}

\begin{figure}
    \centering
    \includegraphics[width=\linewidth]{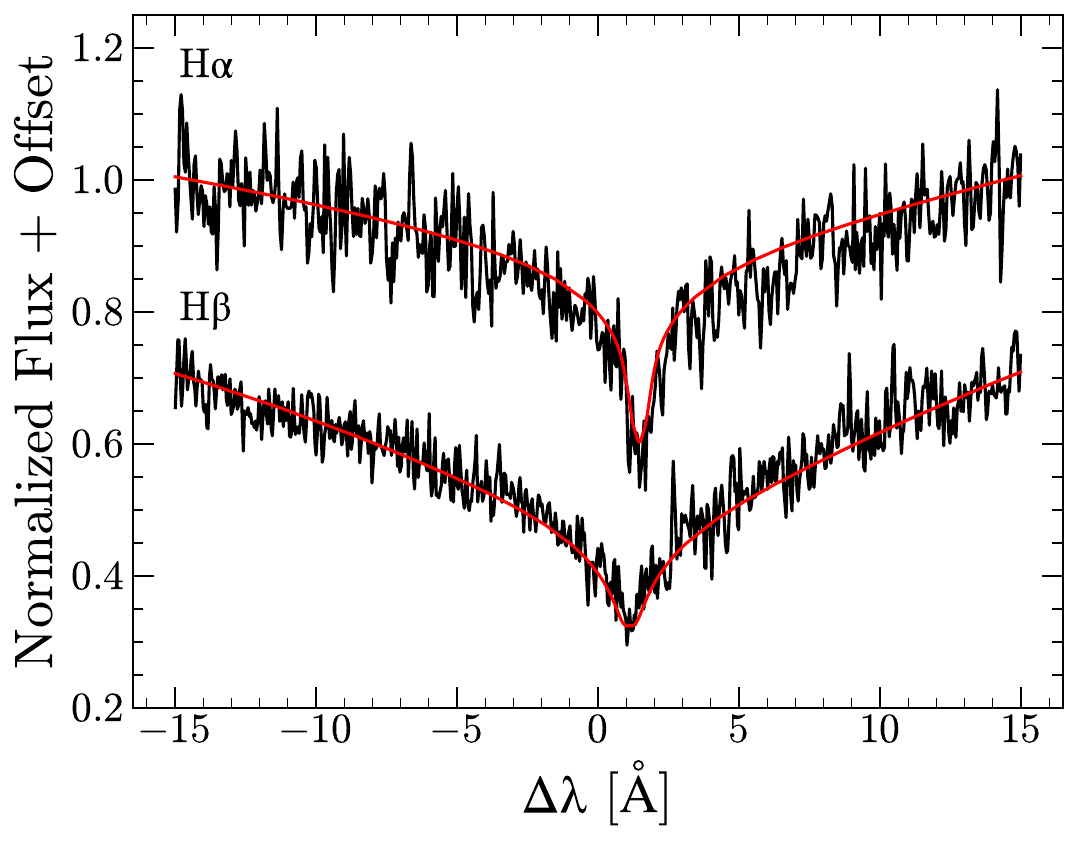}
    \caption{An example \texttt{corv} fit to the H$\alpha$ and H$\beta$ lines of the SPY white dwarf HE 0049-0940 in a window of $\pm 15$~$\angstrom$, yielding a radial velocity of $67.3\pm 2.4$~km s$^{-1}$. SPY spectra have high enough spectral resolution to capture the line cores. Because the redshift is equal in velocity, the lines are linearly offset from one another in wavelength space.}
    \label{fig:nlte}
\end{figure}

\begin{figure*}
    \centering
    \includegraphics[width=\linewidth]{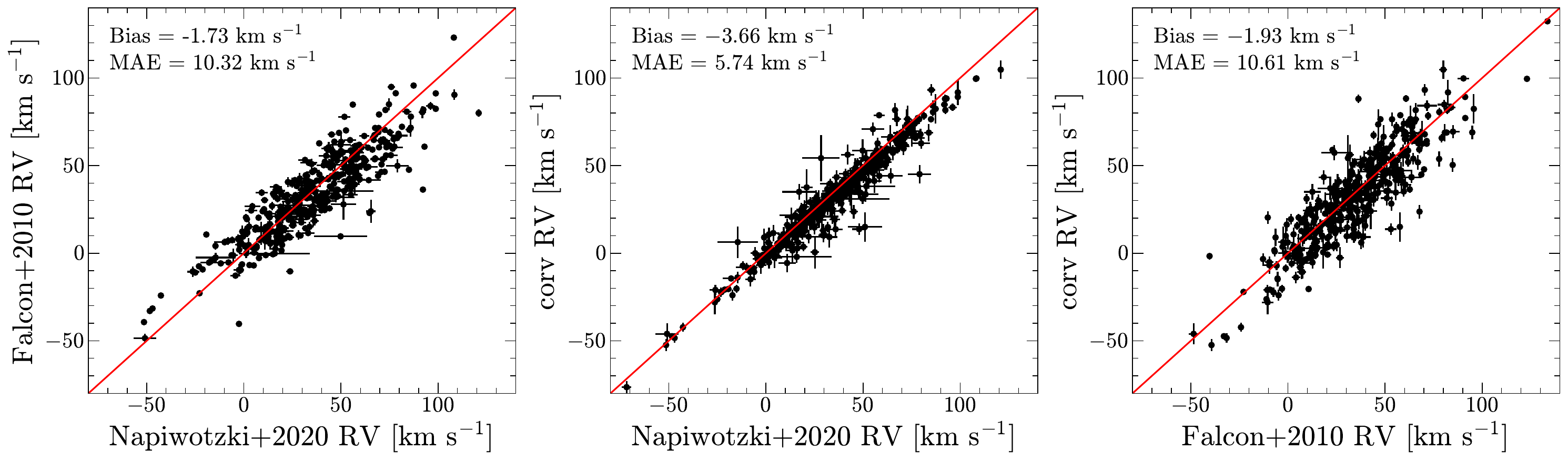}
    \caption{Consistency between the radial velocities of \cite{2020A&A...638A.131N}, \cite{2010ApJ...712..585F}, and this work, with bias and mean absolute error listed. One-to-one lines are marked in red. We find good consistency between the three. The $-3.66$~km s$^{-1}$ bias of our apparent radial velocities relative to \cite{2020A&A...638A.131N} and $-1.93$~km s$^{-1}$ relative to \cite{2010ApJ...712..585F} are on the same scale as the $-1.73$~km s$^{-1}$ bias between the two external analyses.}
    \label{fig:consistency}
\end{figure*}

We measure apparent radial velocities for each individual exposure of each object in the SPY survey by fitting the $\pm 15~\angstrom$ surrounding the H$\alpha$ and H$\beta$ lines, as recommended by \cite{2020A&A...638A.131N} and \cite{2025A&A...695A.131R}, using  \texttt{corv}. This window is optimal to capture redshift information from the line cores.

We fit the 3D DA LTE model spectra to each individual exposure in the SPY sample, convolved to match the spectral resolution of the data. These are high-resolution model spectra which accurately describe the line cores. We ensure that our measurements are accurate by restricting the sample to those objects where \texttt{corv} reports best fit model spectra with $7.01 < \log g < 8.99$ (to remove measurements which fail to converge or pile up on the edge of the parameter space), reduced chi-squared ($\chi^2_r$) less than 5, and radial velocity uncertainty less than $15$~km s$^{-1}$. In addition to removing failed fits, this removes objects that have converged to the edges of the model grid and are therefore unreliable. Finally, we reject all radial velocity measurements which fall three standard deviations outside of the median of the distribution of all measurements which pass the above cuts (the velocity distribution is tight enough that these likely indicate an unphysical measurement). 

Many objects in our sample have multiple exposures. We measure apparent radial velocities for each individual exposure, and then compute the radial velocity for each object as the weighted mean of each exposure. This gives us high-resolution apparent radial velocities for 451 DA white dwarfs and results in a sample mean apparent radial velocity of $31.5$~km s$^{-1}$ and standard deviation of $29.4$~km s$^{-1}$. Figure \ref{fig:nlte} provides an example of one such fit.

To ensure that the radial velocities we measure are precise and to confirm that our treatment of the SPY spectra is correct, we compare our apparent radial velocities against those of \cite{2020A&A...638A.131N} and \cite{2010ApJ...712..585F}. Both these analyses measured the apparent radial velocities of the SPY sample using the line cores. They fit the $\pm 15$~$\angstrom$ about the H$\alpha$ and H$\beta$ lines separately, and then determined the radial velocity of the star as the weighted average of the two lines. The consistency between their analyses and ours is presented in Figure \ref{fig:consistency}. We find good consistency, with a bias of $-3.66$~km s$^{-1}$ relative to the radial velocities of \cite{2020A&A...638A.131N} and $-1.93$~km s$^{-1}$ relative to those of \cite{2010ApJ...712..585F}. This is similar to the bias of \cite{2010ApJ...712..585F} relative to \cite{2020A&A...638A.131N}, which is $-1.73$~km s$^{-1}$.

White dwarfs at effective temperatures associated with ZZ Ceti pulsation have been known to have variable radial velocities \citep{2003ApJ...589..921T}. This is not a significant source of uncertainty in our sample because the exposure time of SPY observations, $1800$~s, is longer than the period of white dwarf pulsations, which have typical periods of $800-1200$~s \citep{2017ApJS..232...23H}. Spectral variations due to pulsations will therefore average out over the course of the exposure.

\subsection{Low-Resolution Apparent Radial Velocities}

\begin{figure*}
    \centering
    \includegraphics[width=\linewidth]{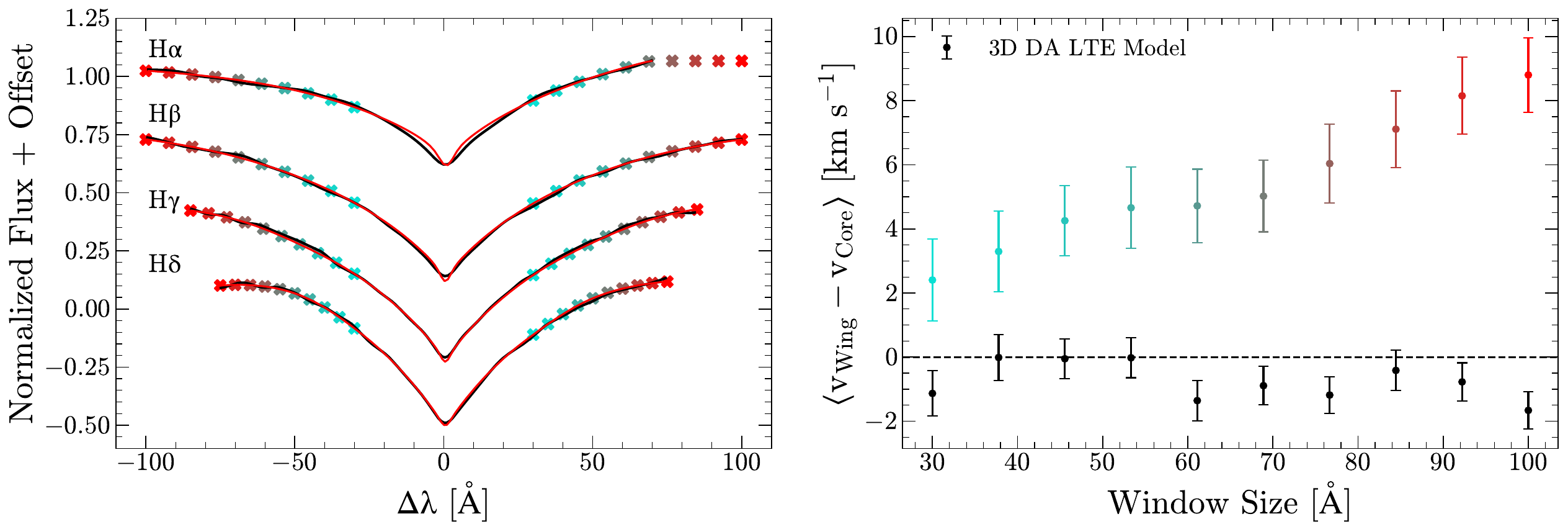}
    \caption{\textit{(Left)} An example \texttt{corv} fit to SPY spectroscopic data ($R = 18,500$) convolved down to approximate SDSS spectral resolution ($R = 1800$). At this resolution, the NLTE core is not distinguishable. Each of the 10 tested fit windows are marked with colors corresponding to their window size. \textit{(Right)} Median resolution-induced radial velocity as a function of window size fit. Actual measurements from the SPY dataset are plotted in colors corresponding to window size. Black is a baseline comparison of resolution-induced radial velocity measurements from model spectra fitted to other model spectra. We find a substantial difference between these datasets, indicating that resolution-induced radial velocity offsets from core to wing fits are present in the SPY data but not the models.}
    \label{fig:window}
\end{figure*}

We measure the apparent radial velocities of the Stark-broadened wings by convolving the high-resolution UVES spectra to $R = 1800$ ($3.6$~$\angstrom$ at H$\alpha$). This is roughly the spectral resolution of the SDSS BOSS spectrograph. At this resolution, the line cores cannot be resolved and radial velocity information is dominated by the wings. 

First, we measure radial velocities from the H$\alpha$, H$\beta$, H$\gamma$, and H$\delta$ lines simultaneously. We fit each individual exposure using 10 different window sizes, ranging from $\pm 30$~$\angstrom$ about all four lines at the narrowest to $\pm 100$~$\angstrom$ at H$\alpha$ and H$\beta$, $\pm 85$~$\angstrom$ at H$\gamma$, and $\pm 75$~$\angstrom$ at H$\delta$ at the widest, in evenly spaced steps. We do this to assess the magnitude of resolution-induced redshifts as a function of window size. Next, in order to assess the magnitude of the effect at individual lines, we fit each exposure using the H$\alpha$, H$\beta$, H$\gamma$, and H$\delta$ lines separately, as well as the H$\alpha$ and H$\beta$ lines simultaneously. We do this using the widest windows of $\pm 100$~$\angstrom$ at H$\alpha$ and H$\beta$, $\pm 85$~$\angstrom$ at H$\gamma$, and $\pm 75$~$\angstrom$ at H$\delta$. As with the high-resolution radial velocities, we ensure that our fits are reliable by requiring that they have $7.01 < \log g < 8.99$, $\chi^2_r < 5$, have radial velocity uncertainty less than $15$~km s$^{-1}$, and are within three standard deviations of the median.

% Occupation probability discussion was here!!

We measure the magnitude of resolution-induced redshifts for each individual exposure as the difference between the apparent radial velocity from the wing-dominated low-resolution spectrum and that of the core-dominated high-resolution spectrum. We then take the weighted average difference across all exposures of each object to determine a resolution-induced redshift for each star in the sample. In total we make reliable measurements for 338 DA white dwarfs using the H$\alpha$, H$\beta$, H$\gamma$, and H$\delta$ lines simultaneously.

\begin{figure*}
    \centering
    \includegraphics[width=\linewidth]{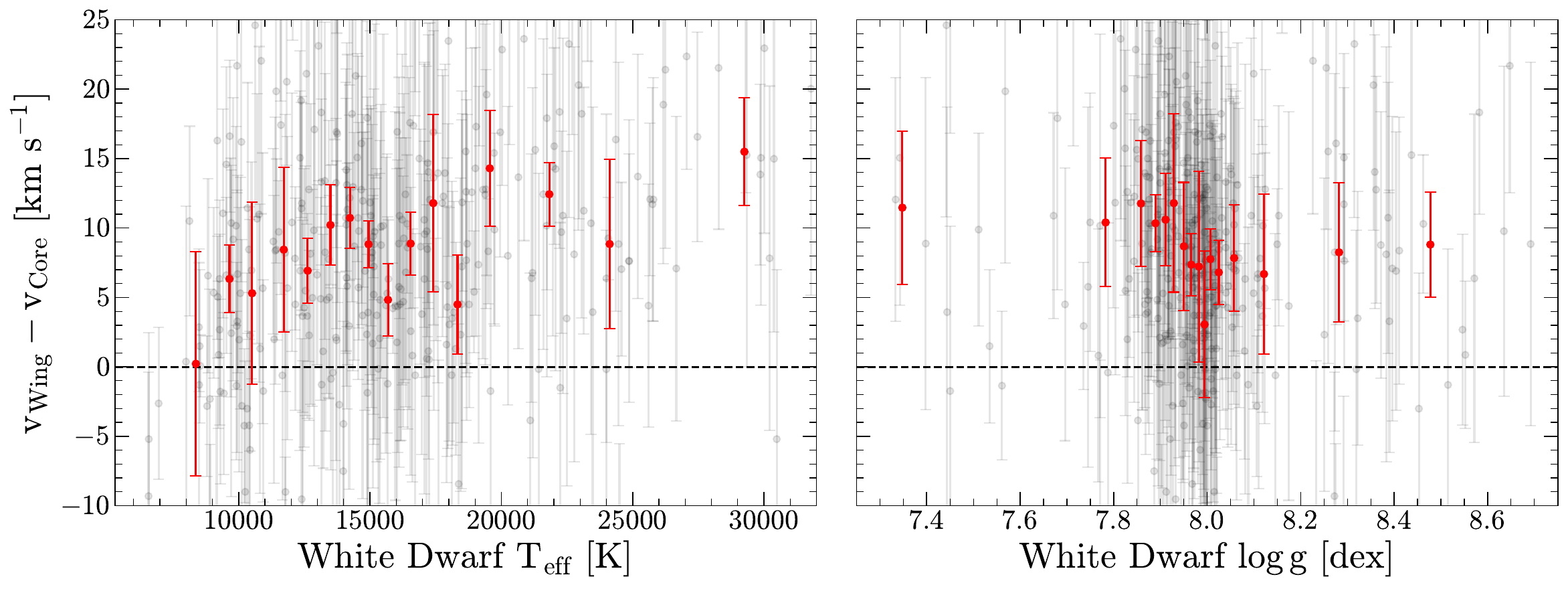}
    \caption{{\it(Left)} Resolution-induced redshifts directly measured from high-resolution SPY spectroscopy as a function of effective temperature. {\it(Right)} The same, but as a function of surface gravity. Binned median differences are plotted in red, with 28 points per bin, over the scatter of individual data points. We find a statistically significant offset between apparent radial velocities measured from the wings and those measured from the line cores. The offset becomes stronger with increasing temperature, but no trend is apparent for surface gravity. We adopt $T_\text{eff}$ and $\log g$ values from \cite{2021MNRAS.508.3877G}.}
    \label{fig:direct}
\end{figure*}

\begin{figure*}
    \centering
    \includegraphics[width=\linewidth]{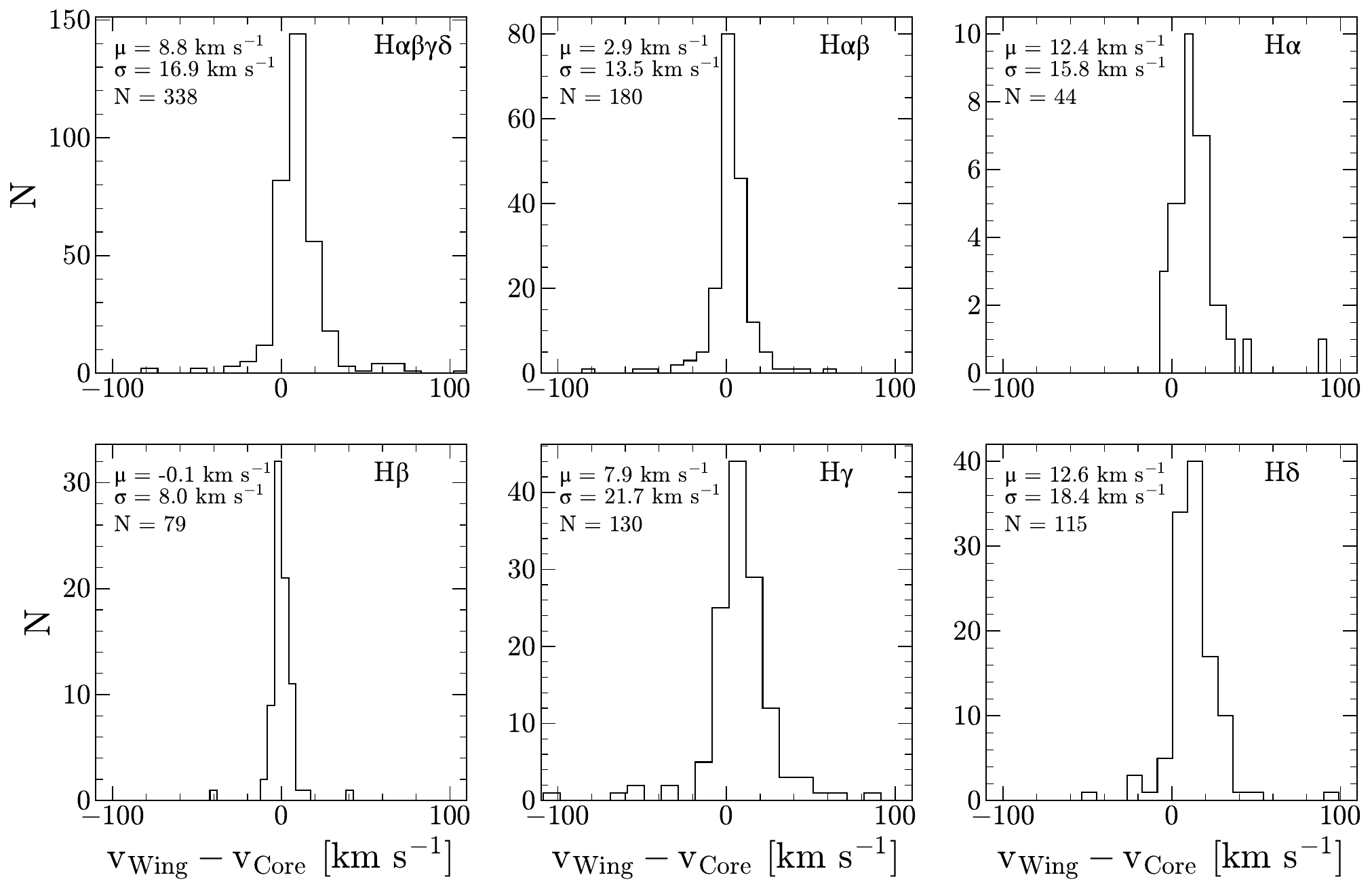}
    \caption{Distribution of measured difference between wing and core velocities for different combinations of lines fitted. We test simultaneously fitting H$\alpha$, H$\beta$, H$\gamma$, and H$\delta$; fitting H$\alpha$ and H$\beta$; as well as all four lines individually. The median, standard deviation, and sample size of each distribution are reported. With the exception of H$\alpha$, magnitude of the offset increases with increasing line order from $-0.1$~km s$^{-1}$ for H$\beta$ to $12.6$~km s$^{-1}$ for H$\delta$.}
    \label{fig:single-line}
\end{figure*}

Figure \ref{fig:window} presents the median resolution-induced redshift as a function of fit window size, as well as an example of the wing fit. To ensure that the measured resolution-induced redshift is actually present in the data and not simply error introduced by \texttt{corv}'s fitting routine, we include a baseline comparison sample. We simulate 300 observations by adding noise to our model spectra, bringing them to a signal-to-noise ratio of $35$ and measure their radial velocities. The median offset of the SPY sample is significantly different from the comparison dataset, indicating that the resolution-induced redshift is actually present in the SPY data. Fitting the wings at all window sizes induces resolution-induced redshifts, and the magnitude of the redshift increases with window size.

We find a consistent and statistically significant offset between wing and core radial velocities as a function of stellar parameters, presented in Figure \ref{fig:direct}. Some temperature dependence is detected, but we are generally unable to resolve its detailed structure due to the scatter of the sample. No dependence on $\log g$ is detected. We note that measuring low-resolution radial velocities using a grid of 1D NLTE model atmospheres \citep{2006ApJ...651L.137K, 2009ApJ...696.1755T, 2011ApJ...730..128T} results in similar resolution-induced redshifts, indicating that this result is not model-dependent. 

Figure \ref{fig:single-line} presents the distribution of the measured resolution-induced redshift for different combinations of fitted lines. We expect that bluer lines should exhibit greater resolution redshifts because equal perturbations in wavelength space will produce greater perturbations in velocity space. Our results are broadly consistent with this expectation: except for the single-line H$\alpha$ distribution, which has the largest resolution-induced redshift, fits which contain more information from higher-order lines tend to have stronger resolution-induced redshifts. This indicates that models of the H$\alpha$ line are likely more sensitive to physics which are not accounted for in the models. We note the presence of a small blueshift in the comparison sample, averaging $-0.75\pm 0.59$~km s$^{-1}$. It is possible that this is a bias induced by \texttt{corv}'s continuum normalization process.

%We calculate line bisectors for each exposure in our sample as well as its best fit model. 

\section{Statistical Measurements from Low-Resolution Spectroscopic Data} \label{sec:low-res}

\subsection{Kinematics in the SDSS Value-Added Catalog}

The specifics of the temperature dependence of the resolution-induced redshifts is difficult to determine from the relatively small sample size of the SPY dataset. We determine the relation more precisely by leveraging the large number of low-resolution spectra in SDSS. The temperature dependence of the shift can be isolated by averaging out the random space motions of a large sample of stars \citep{2010ApJ...712..585F}, binned by common effective temperatures. The magnitude of a star's gravitational redshift can be estimated and subtracted out via its photometrically derived $T_\text{eff}$ and $\log g$ parameters and an assumed mass-radius relation. This leaves a radial velocity signal which consists solely of the star's actual motion and any resolution-induced redshift. We take advantage of the large sample of DA white dwarf spectra in SDSS to make this measurement.

First, we estimate theoretical gravitational redshifts for each star in the sample using the  mass-radius relation of \cite{2020ApJ...901...93B}, with $T_\text{eff}$ and $\log g$ measured by \cite{2021MNRAS.508.3877G} by fitting to Gaia photometry and astrometry. We subtract the theoretical values of gravitational redshift from the apparent radial velocity of the spectrum. The VAC contains apparent radial velocities corrected to the local standard of rest and asymmetric drift biases induced by the uneven distribution of sources on the sky \citep{2024ApJ...977..237C}. In this frame, objects in the solar neighborhood of the Galactic disk have isotropically distributed velocities and so averaging over radial velocities isolates resolution-induced redshift. 

\begin{figure}
    \centering
    \includegraphics[width=\linewidth]{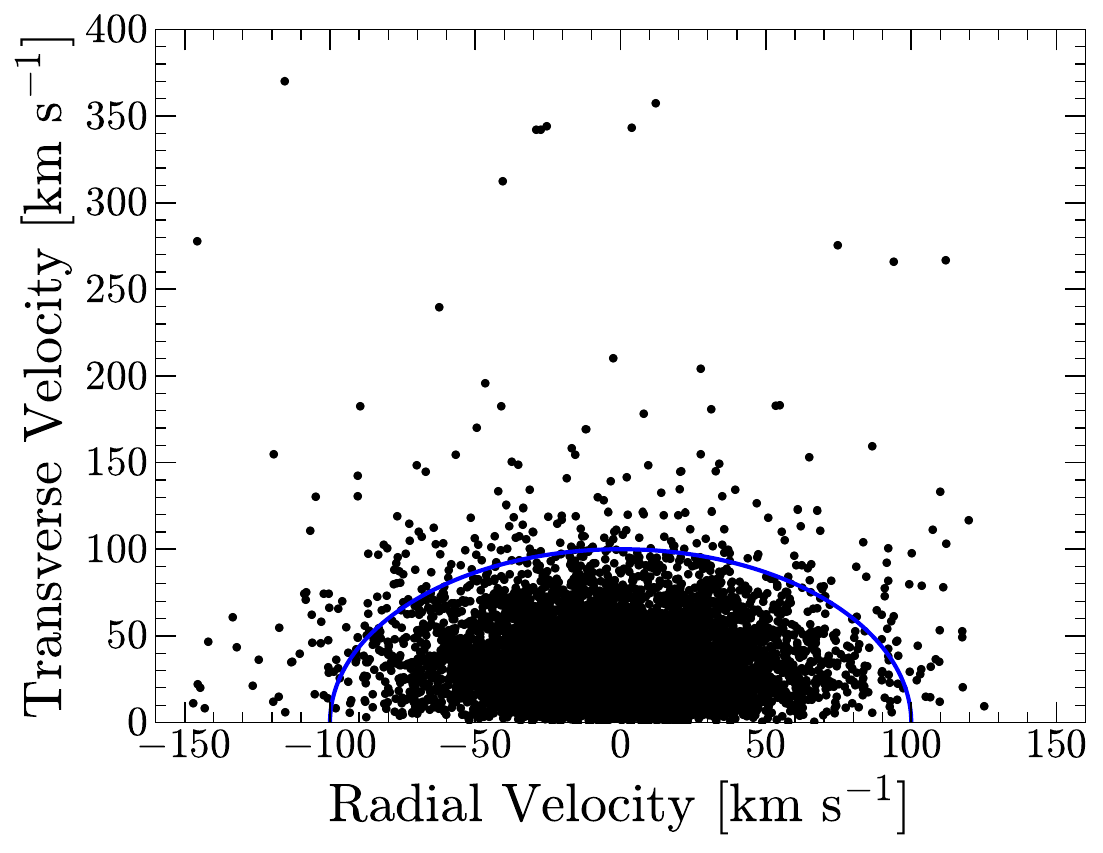}
    \caption{Transverse velocity corrected to the local standard of rest (LSR) versus radial velocity for all stars in our sample. Stars with total velocities greater than $100$~km s$^{-1}$ (marked in blue) are removed from the sample to exclude objects with kinematics inconsistent with the Galactic disk.}
    \label{fig:toomre}
\end{figure}

Averaging out actual motions requires that all stars in the sample have kinematics consistent with the Galactic disk population. We calculate three-dimensional velocities in the frame of the local standard of rest (LSR) for all stars in our sample using their \textit{Gaia} proper motions and median distances inferred by \cite{2021AJ....161..147B}. We adopt the LSR frame measured by \cite{2010MNRAS.403.1829S}:
\begin{equation}
    (U,V,W) = (11.1,12.24, 7.25)\text{~km s}^{-1}    
\end{equation}
Figure \ref{fig:toomre} presents a Toomre diagram of the sample. We require that each object in the sample have total velocity less than $100$~km s$^{-1}$, as recommended by \cite{2022A&A...658A..22R}, and that the entire sample be within $500$~pc so that the variation of velocity dispersion throughout the height of the disk and Galactic rotation are not likely to be significant sources of bias \citep{1989ApJ...342..272F, annurev:/content/journals/10.1146/annurev.aa.27.090189.003011}. Our final sample size is $8022$ white dwarfs, with $1880$ from SDSS-V and $6142$ from previous generations of the survey.

\subsection{Validating the Comoving Approximation}

Accurately measuring the temperature dependence of the resolution-induced redshifts requires choosing temperature bins that are fine enough that structure can be distinguished but coarse enough that the sample size is sufficient to average out random space motions. We choose optimal bins by investigating transverse velocities as a function of apparent radial velocity.

\begin{figure}
    \centering
    \includegraphics[width=\linewidth]{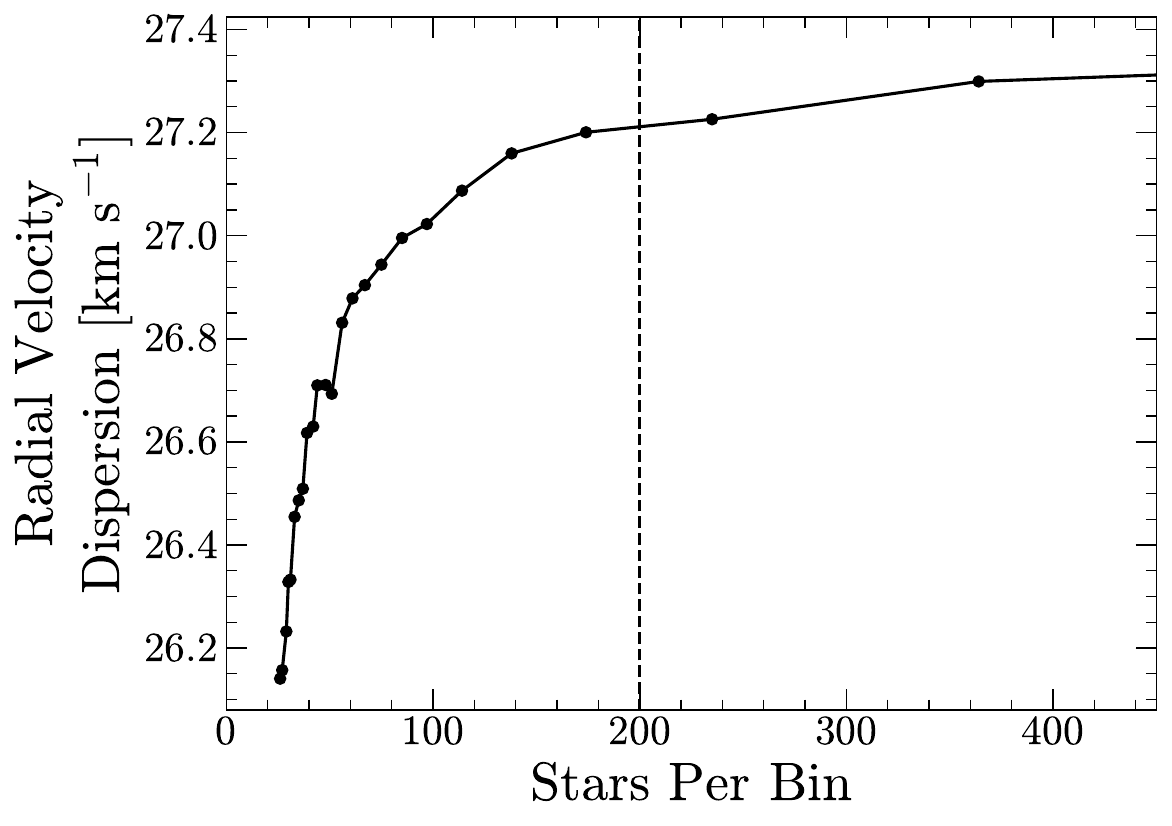}
    \caption{Radial velocity dispersion as a function of stars per bin. We find a knee-like response curve such that averaging becomes poorly behaved with less than 100 stars per bin. Therefore we bin our sample with 200 stars per bin (marked).}
    \label{fig:binnum}
\end{figure}

\begin{figure*}
    \centering
    \includegraphics[width=\linewidth]{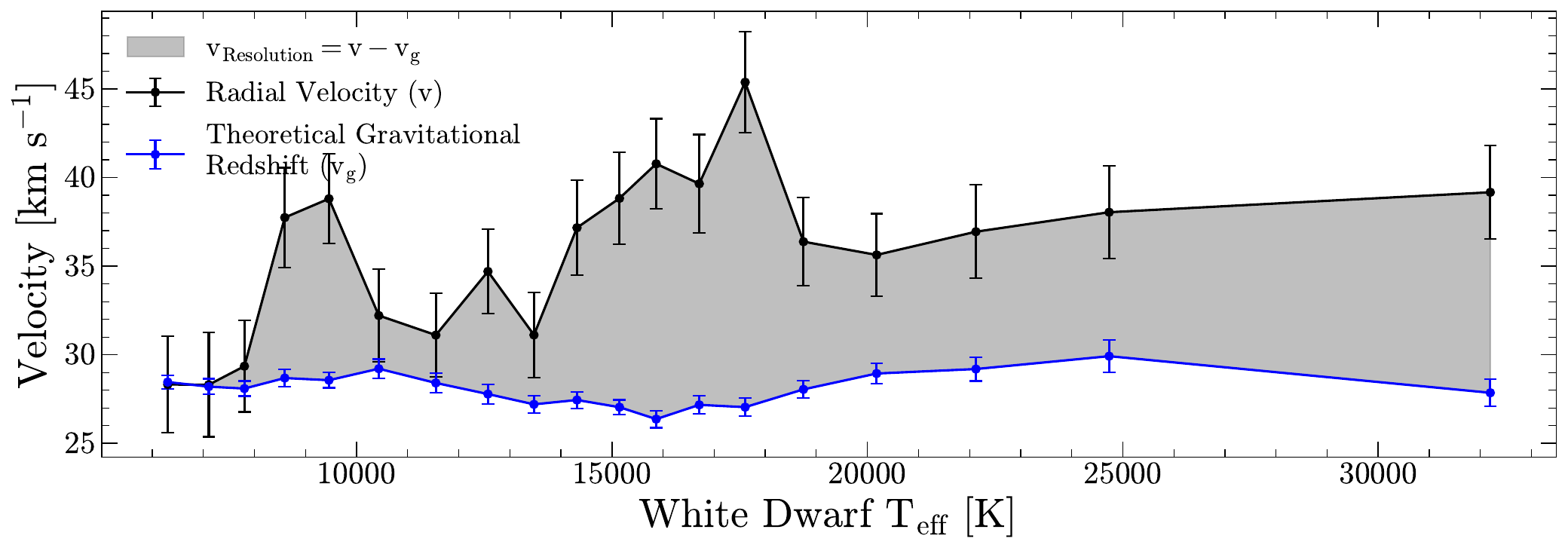}
    \caption{Median apparent radial velocities and theoretical gravitational redshifts of each temperature bin for DAs in SDSS with $0.55~M_\odot \leq M \leq~0.65~M_\odot$. We calculate the resolution-induced radial velocity for each object as the difference between its apparent radial velocity and expected gravitational redshift (the region in gray). Our apparent radial velocities are consistently greater than the theoretical gravitational redshift (using the mass-radius relation of \citealt{2020ApJ...901...93B}) for white dwarfs observed with low-resolution SDSS spectra.}
    \label{fig:define-vstark}
\end{figure*}

We create bins of equal size in transverse velocity using different numbers stars per bin, ranging from approximately 802 (10 bins in total) to approximately 26 (300 bins in total). We calculate the radial velocity dispersion in each bin as the standard deviation of the radial velocities subtracted in quadrature by the mean radial velocity uncertainty. If the sample kinematics are not properly averaged out, then the velocity dispersion will be unphysically small. Figure \ref{fig:binnum} presents the radial velocity dispersion as a function of stars per bin. We find that at least 100 stars per bin are required to reliably average out random space motions. On the basis of this analysis, we bin our data such that each bin contains 200 stars, comfortably allowing us to average out random space motions while still providing good resolution to distinguish potential temperature dependencies. This choice is supported by \cite{2012ApJ...757..116F} who found that such a statistical measurement was possible with as few as 36 stars. 

\cite{2018MNRAS.476.2584M} find that $9.5\pm2.0\%$ of white dwarfs are in binaries with separations $<4$~au. Although some of these stars will be filtered by the $\eta > 2.5$ and $M < 0.35M_\odot$ cut, the majority (66\%) of stars in the SDSS sample have only a single exposure. These stars are less likely to be identified as binaries, so choosing to bin more stars than necessary minimizes the contaminant effects of undetected binary systems.

\begin{figure*}
    \centering
    \includegraphics[width=\linewidth]{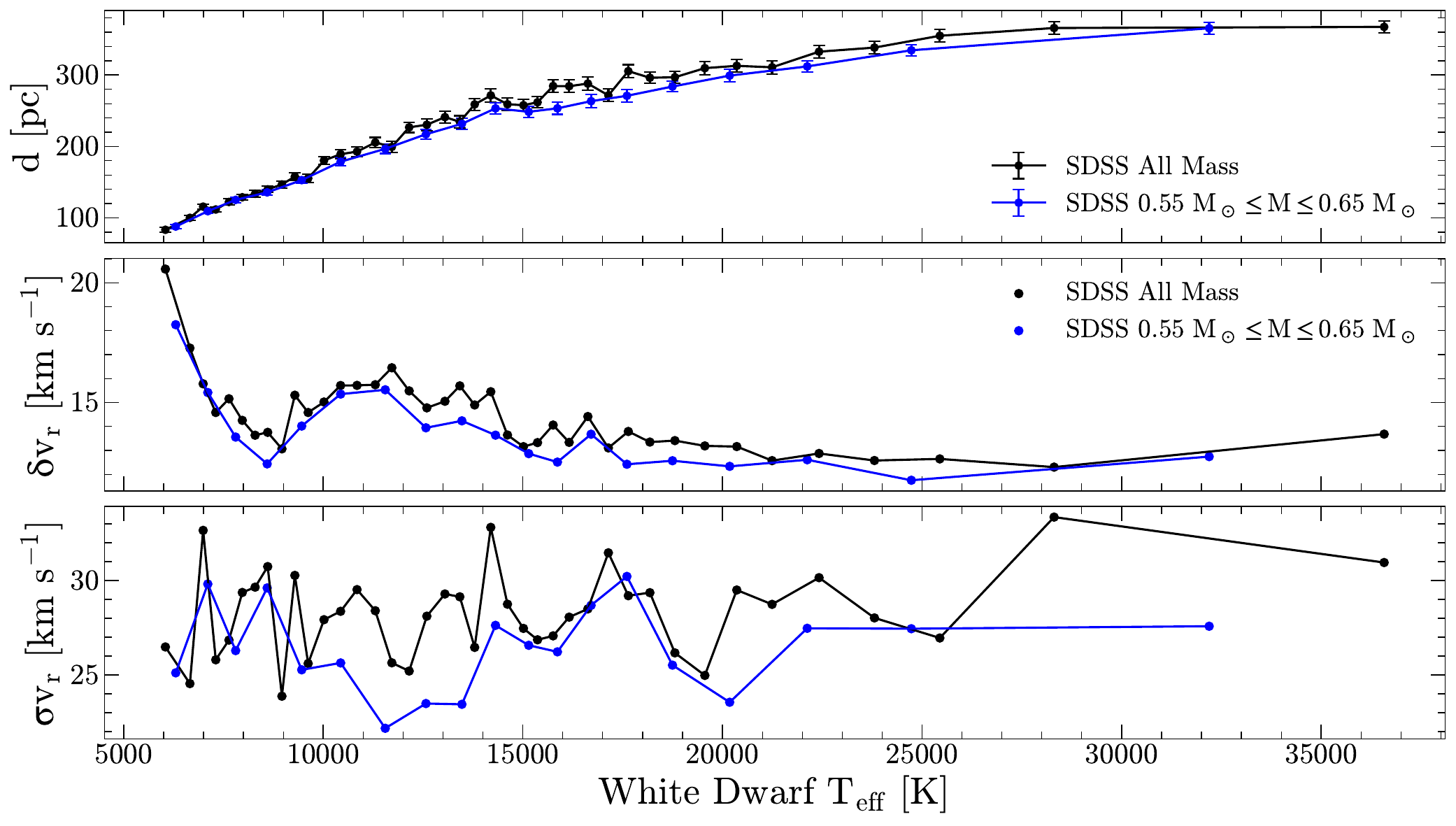}
    \caption{\textit{(Top)} Median distance ($d$) as measured by \cite{2021AJ....161..147B} for the entire SDSS sample in black as well as the mass-restricted subset in blue. \textit{(Middle)} Median uncertainties of radial velocity measurements ($\delta v_r$) within each temperature bin. The median uncertainties for most of the sample are between $12-14$~km s$^{-1}$ except for bins below $\approx7000$~K, where they rise to between $18-20$~km s$^{-1}$. This is because at these temperatures the Balmer lines become weaker, making fitting more challenging. \textit{(Bottom)} Velocity dispersion ($\sigma v_r$) within each bin, calculated by subtracting the median radial velocity uncertainty from the standard deviation of the radial velocity distribution in quadrature. Our results are consistent with the flat age-velocity dispersion of \cite{2022A&A...658A..22R} across all ages.}
    \label{fig:sdsssample}
\end{figure*}

We create an additional sub-sample consisting of white dwarfs with photometrically measured masses in the range $0.55~M_\odot \leq M \leq~0.65~M_\odot$. At equivalent mass and temperature, different white dwarfs will have the same gravitational redshift. Restricting masses further ensures that biases due to gravitational redshifts are not present in the sample. This gives us $3904$ objects in the $0.60\pm0.05~M_\odot$ sample. Figure \ref{fig:define-vstark} presents the median apparent radial velocity and theoretical gravitational redshift using the mass-radius relation of \cite{2020ApJ...901...93B} within each temperature bin for the $0.60\pm0.05~M_\odot$ sample.

Figure \ref{fig:sdsssample} presents the median distance as calculated by \cite{2021AJ....161..147B}, the median uncertainty of radial velocity measurements within each temperature bin, and the velocity dispersion within each bin. The median uncertainties for most of the sample are between $12-14$~km s$^{-1}$ except for bins below $\approx7000$~K, where they rise to between $18-20$~km s$^{-1}$. This is because at these temperatures the Balmer lines become weaker, making fitting more challenging. We calculate the velocity dispersion by subtracting the median radial velocity uncertainty from the standard deviation of the radial velocity distribution in quadrature. For all bins, the median signal-to-noise ratio and $G$-band magnitude with respective standard deviations are $15.5\pm2.7$ and $18.4\pm0.2$~mag the entire sample and $16.6\pm2.7$ and $18.2\pm0.2$~mag for the mass-restricted sample. The hottest white dwarf bins have median distances of up to $350$~pc, but are brighter. Our velocity dispersion is flat, with average values of $28.3$ and $26.4$~km s$^{-1}$ for the full and mass restricted samples respectively. This is consistent with the age-velocity dispersion of \cite{2022A&A...658A..22R}, who also found a flat velocity dispersion relation. We note that \cite{2022A&A...658A..22R} considered velocity dispersion as a function of total age whereas we only consider it as a function of $T_\text{eff}$, which is a proxy for cooling age.

\section{Results} \label{sec:results}

\subsection{Resolution-Induced Redshifts in the SDSS Sample}

\begin{figure*}
    \centering
    \includegraphics[width=\linewidth]{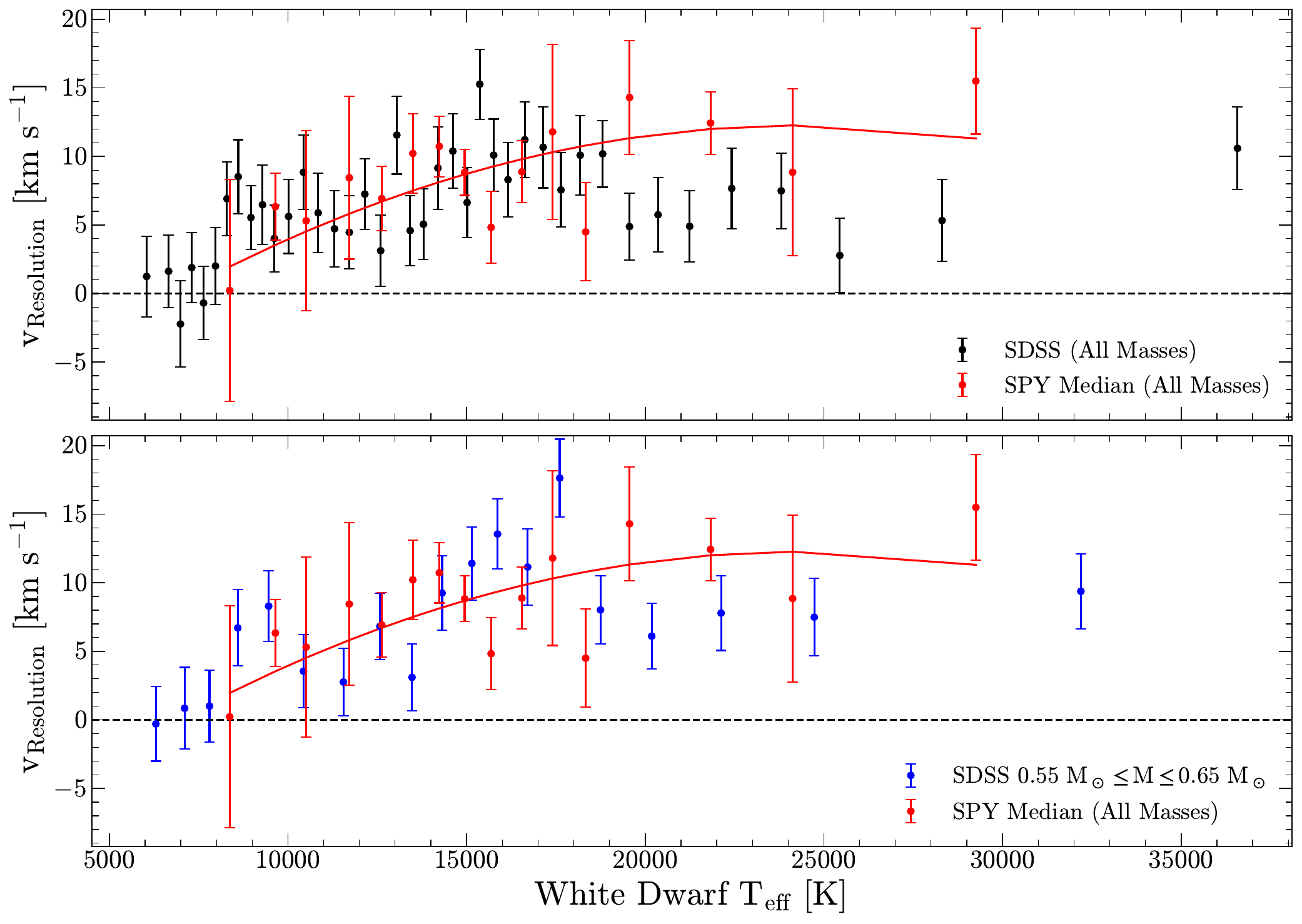}
    \includegraphics[width=\linewidth]{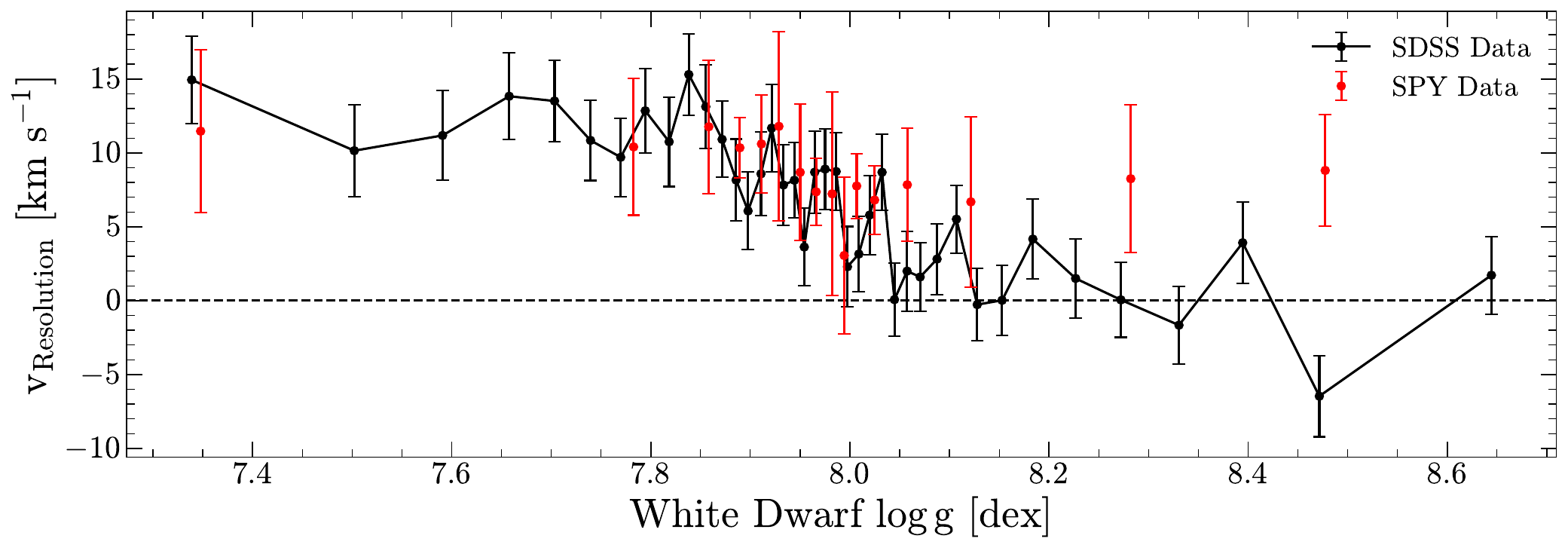}
    \caption{{\it(Top)} Resolution-induced redshift (difference in observed redshift versus theoretically predicted gravitational redshift, see Figure \ref{fig:define-vstark}) as a function of temperature for the entire SDSS sample in black. Random space motions average out within each bin, leaving only the resolution-induced redshift. The resolution-induced redshifts from the SPY sample measured via model atmospheres are plotted in red. The best fit line is also plotted in red for the temperature range of the SPY data. {\it(Middle)} The same, but for a restricted mass sample with $0.60\pm0.05~M_\odot$. The restricted mass sample has standard deviation in expected gravitational redshift of $6.4$~km s$^{-1}$. {\it(Bottom)} Resolution-induced redshift as a function of photometric surface gravity. We detect a dependence on surface gravity in SDSS data, with white dwarfs with low surface gravity exhibiting greater resolution-induced biases. No such dependence is detectable with SPY. This may be because the distribution of surface gravities is sharply peaked around $\log g\approx 8$, so that the SPY data do not contain enough observations of white dwarfs with extreme surface gravity to confidently identify a dependence.}
    \label{fig:statistical}
\end{figure*}

Figure \ref{fig:statistical} presents measured resolution-induced redshift as a function of effective temperature for both SDSS samples and the median resolution-induced redshift measured directly from the SPY dataset. Additionally, it presents the measured resolution-induced redshift as a function of photometrically measured surface gravity from \cite{2021MNRAS.508.3877G}. Generally spectroscopic surface gravities are considered more accurate than photometric surface gravities. Despite this, we adopt photometric surface gravities as they are less dependent on model Balmer line shapes.

The redshifts in the SDSS samples are generally consistent with that of the SPY sample for both effective temperature and surface gravity. We note that redshifts become significant at temperatures greater than $\approx 8,000$~K. This is consistent with the results of \cite{1992ApJ...387..288B}, who found that model spectra are sensitive to the effects of photospheric convection within the range $8000~\text{K} < T_\text{eff} < 15,000~\text{K}$, with a maximum sensitivity at $T_\text{eff} = 13,000~\text{K}$. Because the low-resolution radial velocity measurements probe deeper into the photosphere, it is likely that they are more substantially impacted by convection. Redshifts from SDSS show distinct dependence with surface gravity. Low-mass white dwarfs exhibit larger resolution-induced redshifts compared to massive white dwarfs which exhibit redshifts consistent with zero. The sharply peaked distribution of surface gravities and the limited size of SPY means that surface gravity dependence cannot be resolved via direct measurements.

%We note the presence of one additional peak in the $0.60\pm0.05~M_\odot$ sample associated with the high-temperature ($\approx 24,000$~K) bins. At equal masses hotter white dwarfs are brighter, meaning that these bins of high effective temperature are biased toward more distant stars. At the highest temperatures median distances reach up to $350$~pc. At these distances, random motions may not average out as reliably due to differential rotation in the Galactic disc.

Energy levels which are strongly split by the Stark effect become more easily dissociated by Coulomb interactions with surrounding ions, which can cause the $n=2$ hydrogen state to become inaccessible. State of the art model atmosphere codes model this inaccessibility by means of a calibrated occupation probability \citep{2009ApJ...696.1755T}, treating the probability that the dissociated state can be occupied as a calibrated free parameter. The occupation probability parameterization remains in common use, and is included in the model atmospheres of \cite{2010MmSAI..81..921K} and \cite{2013A&A...559A.104T, 2015ApJ...809..148T}.

Measurements relying on the outer wings of absorption lines may be affected by model parameterizations such as occupation probabilities. We test the effect of model choice on our measurements by comparing apparent radial velocities measured from the SPY spectra convolved to low spectral resolution via fitting model atmospheres to those measured via fitting analytic Voigt profiles. Voigt profiles are more flexible than model atmospheres and there are no inherent assumptions about occupation probabilities; therefore Voigt profiles are free to take the optimal shape of the observed data. We fit the entire SPY dataset convolved to low-resolution using Voigt profiles and require that the fits converge with $\chi^2_r < 2$. We find good agreement between the radial velocities measured with Voigt profiles and model atmospheres, with mean absolute error of $6.62$~km s$^{-1}$ and bias of $1.19$~km s$^{-1}$.

\subsection{Best Practices With SDSS Data} \label{sec:practices}

Because the offset between BOSS radial velocities from SDSS-V and legacy versions of the survey are of similar magnitude and opposite direction to the measured resolution-induced redshifts, it is tempting to think that they are due to a wavelength miscalibration in SDSS-V.\footnote{\cite{2012ApJ...749L..11B} found a mean gravitational redshift of $31.0\pm0.4$~km s$^{-1}$ from radial velocity measurements of legacy BOSS spectra. This measurement is consistent with expectations given the mean mass of the sample, but was likely a fortunate coincidence arising from the cancellation of the $7.1\pm0.7$~km s$^{-1}$ systematic blueshift in the early SDSS wavelength solution and the $8.8$~km s$^{-1}$ median resolution-induced redshift.} However, this is very unlikely because the resolution-induced redshifts observed from SDSS-V are consistent with those of SPY data. The agreement between these separate analyses of independent datasets suggests that accurate white dwarf radial velocities should be measured from SDSS spectra by relying on data from SDSS-V where possible, and by adding a redshift of $7.1\pm0.7$~km s$^{-1}$ (the weighted mean offset) to legacy spectra when SDSS-V data are not available. 

Our measurements of resolution-induced redshifts in the SPY sample are independent of assumptions of mass-radius relation, meaning that they can be used to correct this effect in the SDSS sample. We first fit constant, linear, quadratic, cubic, and quartic polynomials to the offsets from the SPY data. The preferred model of resolution-induced radial velocity ($v_\text{Resolution}$) in km s$^{-1}$ as a function of effective temperature in $K$ is the linear expression (with $\chi^2_r = 0.93$):
\begin{align}\label{equation:one}
% -4.02163235e-08,  1.96083600e-03, -1.16353345e+01
    v_\text{Resolution} = (-4.0\times10^{-8})T_\text{eff}^2 + (1.9\times10^{-3})T_\text{eff} - 12
\end{align}
as displayed in Figure \ref{fig:statistical}. We report similar best-fit polynomials for a range of spectral resolutions in Table \ref{tab:correction}. We find that corrections are necessary at all resolutions, and so we recommend fitting using only line cores when spectral resolution is high enough to resolve them.

We apply a correction to the measured SDSS radial velocities within the temperature range of the SPY data, $8600~K < T_\text{eff} < 27,600~K$, by subtracting out the resolution-induced radial velocity from the above equation. We then divide the dataset into bins of equal mass measured from the spectral energy distribution, with 200 stars per bin, and compute the median corrected and uncorrected radial velocity in each bin. Having subtracted out bias, the median radial velocity in each bin consists solely of gravitational redshift. 

\begin{deluxetable}{ccccc}
\label{tab:params}
\tablecaption{Median effective temperature ($T_\text{eff}$), $\log g$, and observed gravitational redshift with debiasing corrections applied ($v_\text{g,obs}$) for the entire SDSS sample and the mass-restricted sample. The last column presents the theoretically predicted gravitational redshift ($v_\text{g,th}$) for the median temperature and $\log g$ in each sample from the hydrogen-thick mass-radius relations of \cite{2020ApJ...901...93B}.}
\tablewidth{1300pt}
\tabletypesize{\scriptsize}
\tablehead{Sample & \colhead{$T_\text{eff}$ $\left[K\right]$} & \colhead{$\log g$ [dex]} & \colhead{$v_\text{g,obs}$ [km s$^{-1}$]} & \colhead{$v_\text{g,th}$ [km s$^{-1}$]}}
\startdata
\text{All Masses} & 15,356 & 7.97 & $32.0 \pm 0.5$ & 28.7 \\ 
$0.60\pm0.05$~$M_\odot$ & 15,855 & 7.96 & $29.0\pm0.7$ & 28.3 \\ 
\enddata
\end{deluxetable}

\begin{figure*}
    \centering
    \includegraphics[width=\linewidth]{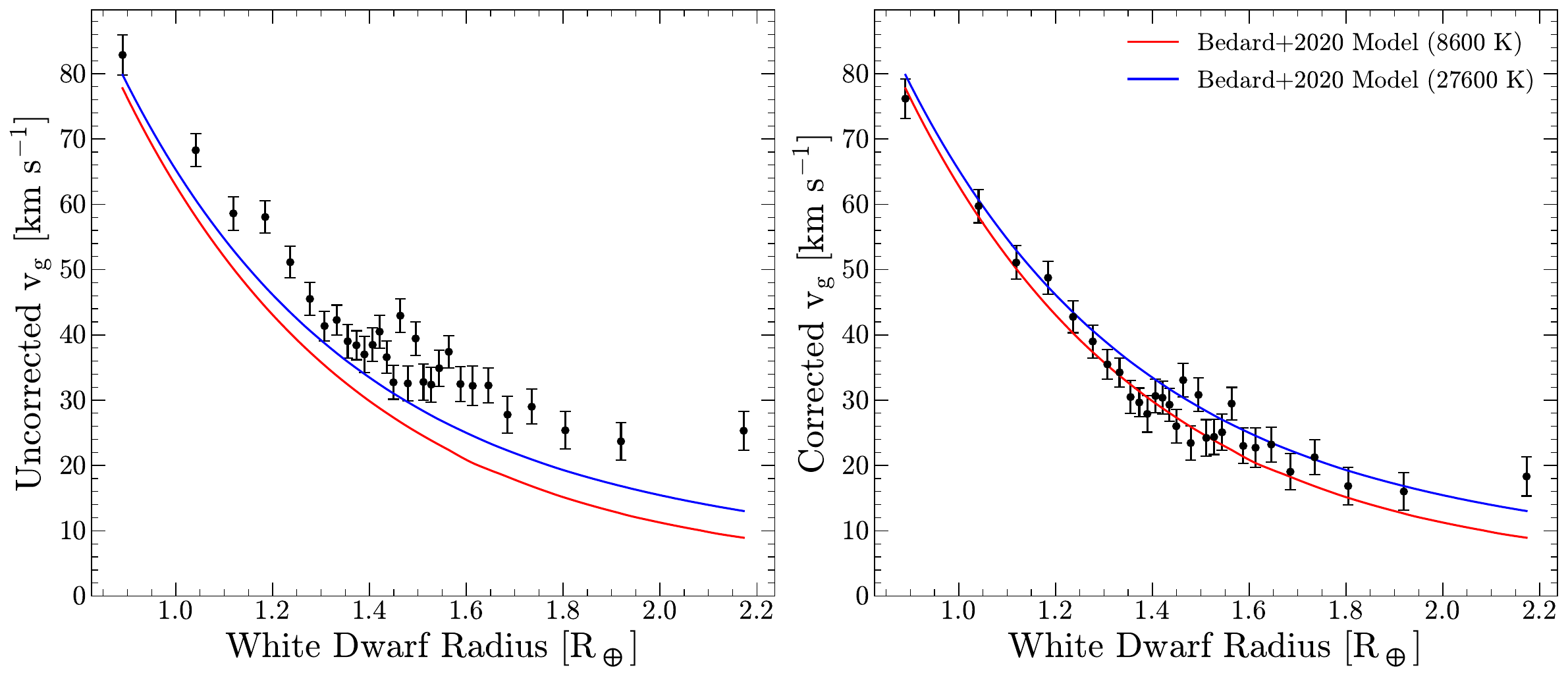}
    \caption{The radius-gravitational redshift relation with and without corrections for resolution-induced redshift from SPY observations (as provided by Equation \ref{equation:one}) compared to theoretical models from \cite{2020ApJ...901...93B}. Data points show the mean radius measurements inferred from photometric fits by \cite{2021MNRAS.508.3877G} using the mass-radius relation of \cite{2020ApJ...901...93B} within each bin. Temperatures plotted correspond to the minimum and maximum temperatures of the SPY data, such that the data should lie between the two curves. We find that measuring mean gravitational redshift from radial velocities corrected for resolution-induced redshift and for offsets in eSDSS relative to SDSS-V yields much better agreement with theoretical models.}
    \label{fig:mass-radius}
\end{figure*}

We present the measured relationship between radius and gravitational redshift in Figure \ref{fig:mass-radius}, using redshift values reported from photometric fits in the VAC, as well as theoretical radius-gravitational redshift relations from \cite{2020ApJ...901...93B} for white dwarfs with canonically-thick hydrogen layers of $M_H/M_\text{WD} = 10^{-4}$. Measurement error induces a bias which flattens the sharply peaked white dwarf radius distribution. We account for this bias via the Monte Carlo method described in \cite{2024ApJ...977..237C}. Applying corrections for resolution-induced redshifts results in much better agreement with the theoretical relation. Table \ref{tab:params} presents the median temperature, $\log g$, and gravitational redshift for both samples, as well as the theoretically expected median gravitational redshift from the mass-radius relation of \cite{2020ApJ...901...93B}.

\section{Summary and Conclusions} \label{sec:discuss}

We have presented a measurement of resolution-induced redshifts in fits to DA white dwarf spectra using two independent datasets. These redshifts exhibit some variation as a function of stellar parameters. They tend to increase in magnitude with increasing effective temperature and decrease in magnitude with increasing surface gravity. Because the wings of hydrogen absorption lines form deeper in the stellar photosphere in regions of high pressure and temperature, this result is consistent with the conclusion that these redshifts result from higher-order physics which are not included in state-of-the-art model spectra. Missing physics potentially includes higher-order Taylor expansions of the Stark effect and non-dipole electrodynamic interactions between charged particles in the stellar photosphere.  

In addition, we characterize an offset in radial velocity measurements made from BOSS spectra in SDSS-V relative to those of previous generations of the spectra. By comparing SDSS radial velocity biases to those of the independent SPY dataset, we show that SDSS-V radial velocities are better calibrated, especially in the blue arm of the spectrograph. We suggest applying a $+7.1\pm0.7$~km s$^{-1}$ correction for radial velocities from legacy BOSS spectra to mitigate this calibration issue.

Because our measurement relies solely on velocities measured using different parts of lines, the strength of this effect using the SPY sample is independent of any assumptions of mass-radius relation. Therefore these measurements can be used to correct the radial velocities of low-resolution samples such as those of \cite{2020ApJ...899..146C}, \cite{2024ApJ...977..237C}, and \cite{2024ApJ...963...17A} to better understand white dwarf stellar structure and probe new areas of white dwarf stellar physics. Those studies measured gravitational redshifts mostly from legacy SDSS spectra, meaning that the effect of resolution-induced redshifts are partially mitigated by relative blueshift between legacy spectra and SDSS-V. It is likely however that they still exhibit resolution-induced biases on the order of $1$~km s$^{-1}$. 

We intend to apply these corrections to measure the mean mass of the white dwarf hydrogen layer using large samples of SDSS and DESI spectroscopy. Models assuming a thick hydrogen layer exhibit gravitational redshift differences on the order of $1$~km s$^{-1}$ compared to those which assume thin hydrogen layers. Therefore resolution-induced redshift corrections are critical to accurately make this measurement. Our work provides a set of best practices for leveraging the vast sample size of low-resolution white dwarf spectroscopic data acquired by all-sky surveys.

\section*{Acknowledgements}

We thank the anonymous referee for their constructive feedback, which improved the quality of this manuscript.
We acknowledge useful insights from Bart Dunlap, Mike Montgomery, and Thomas Gomez regarding the computation of Balmer absorption line shapes, as well as Roberto Raddi. N.R.C is supported by the National Science Foundation Graduate Research Fellowship Program under Grant No. DGE2139757. Any opinions, findings, and conclusions or recommendations expressed in this material are those of the author and do not necessarily reflect the views of the National Science Foundation. N.L.Z. acknowledges visit support from the Institute for Advanced Study (Princeton, NJ). G.A.P. is supported in part by the JHU President's Frontier Award to N.L.Z.

Funding for the Sloan Digital Sky Survey V has been provided by the Alfred P. Sloan Foundation, the Heising-Simons Foundation, the National Science Foundation, and the Participating Institutions. SDSS acknowledges support and resources from the Center for High-Performance Computing at the University of Utah. SDSS telescopes are located at Apache Point Observatory, funded by the Astrophysical Research Consortium and operated by New Mexico State University, and at Las Campanas Observatory, operated by the Carnegie Institution for Science. The SDSS web site is \url{www.sdss.org}.

SDSS is managed by the Astrophysical Research Consortium for the Participating Institutions of the SDSS Collaboration, including the Carnegie Institution for Science, Chilean National Time Allocation Committee (CNTAC) ratified researchers, Caltech, the Gotham Participation Group, Harvard University, Heidelberg University, The Flatiron Institute, The Johns Hopkins University, L'Ecole polytechnique f\'{e}d\'{e}rale de Lausanne (EPFL), Leibniz-Institut f\"{u}r Astrophysik Potsdam (AIP), Max-Planck-Institut f\"{u}r Astronomie (MPIA Heidelberg), Max-Planck-Institut f\"{u}r Extraterrestrische Physik (MPE), Nanjing University, National Astronomical Observatories of China (NAOC), New Mexico State University, The Ohio State University, Pennsylvania State University, Smithsonian Astrophysical Observatory, Space Telescope Science Institute (STScI), the Stellar Astrophysics Participation Group, Universidad Nacional Aut\'{o}noma de M\'{e}xico, University of Arizona, University of Colorado Boulder, University of Illinois at Urbana-Champaign, University of Toronto, University of Utah, University of Virginia, Yale University, and Yunnan University.

% Support for NPR comes from the John D. and Catherine T. MacArthur foundation. 

This work has made use of data from the European Space Agency (ESA) mission \textit{Gaia} (\url{https://www.cosmos.esa.int/gaia}), processed by the \textit{Gaia} Data Processing and Analysis Consortium (DPAC, \url{https://www.cosmos.esa.int/web/gaia/dpac/consortium}). Funding for the DPAC has been provided by national institutions, in particular the institutions participating in the \textit{Gaia} Multilateral Agreement.

Based on data obtained from the ESO Science Archive Facility with DOI(s): \url{https://doi.eso.org/10.18727/archive/50}.

This research has made use of the VizieR catalogue access tool, CDS,
Strasbourg, France \citep{10.26093/cds/vizier}. The original description 
of the VizieR service was published in \citet{vizier2000}.

\bibliography{citations.bib}{}
\bibliographystyle{aasjournal}

\appendix \label{app:a}
\restartappendixnumbering
\section{Correction Values For Different Resolutions}

We provide polynomial best fits for radial velocity corrections at different spectral resolutions. Results are provided in the form:
\begin{equation}
    v_\text{Resolution} = \sum_{i=0}^n a_i T_\text{eff}^i.
\end{equation}
Additionally, we report the reduced chi-squared statistic $\chi^2_r$ of the polynomial fit relative to the SPY data. Models were chosen based on the polynomial degree with reduced chi-squared closest to one.

\begin{deluxetable}{cccccccc}[h!]
\label{tab:correction}
\tablecaption{Correction Values For Different Resolutions}
\tablewidth{1300pt}
\tabletypesize{\scriptsize}
\tablehead{\colhead{R} & \colhead{$a_0$} & \colhead{$a_1$} & \colhead{$a_2$} & \colhead{$a_3$} & \colhead{$a_4$} & \colhead{$a_5$} & \colhead{$\chi^2_r$}}
\startdata
1800 & -12 & $1.9\times 10^{-3}$ & $-4.0\times10^{-8}$ & - & - & - & 0.93 \\
2000 & $-3.71\times10^{2}$ & $1.21\times10^{-1}$ & $-1.54\times10^{-5}$ & $9.63\times10^{-10}$ & $-2.90\times10^{-14}$ & $3.35\times10^{-19}$ & $3.13$ \\
3000 & $9.75\times10^{0}$ & - & - & - & - & - & $1.00$ \\
4000 & $-1.59\times10^{2}$ & $4.59\times10^{-2}$ & $-4.82\times10^{-6}$ & $2.39\times10^{-10}$ & $-5.55\times10^{-15}$ & $4.84\times10^{-20}$ & $0.92$ \\
5000 & $1.05\times10^{0}$ & $4.58\times10^{-4}$ & $2.21\times10^{-9}$ & - & - & - & $0.65$ \\
6000 & $-2.47\times10^{2}$ & $7.34\times10^{-2}$ & $-8.10\times10^{-6}$ & $4.30\times10^{-10}$ & $-1.09\times10^{-14}$ & $1.06\times10^{-19}$ & $0.74$ \\
7000 & $-2.95\times10^{2}$ & $8.78\times10^{-2}$ & $-9.62\times10^{-6}$ & $5.02\times10^{-10}$ & $-1.25\times10^{-14}$ & $1.19\times10^{-19}$ & $0.97$ \\
8000 & $-1.51\times10^{2}$ & $4.99\times10^{-2}$ & $-5.80\times10^{-6}$ & $3.16\times10^{-10}$ & $-8.15\times10^{-15}$ & $7.98\times10^{-20}$ & $1.13$ \\
9000 & $1.22\times10^{1}$ & - & - & - & - & - & $1.08$ \\
10000 & $1.15\times10^{1}$ & - & - & - & - & - & $1.22$ \\
11000 & $2.18\times10^{1}$ & $-3.16\times10^{-3}$ & $2.61\times10^{-7}$ & $-8.73\times10^{-12}$ & $1.07\times10^{-16}$ & - & $0.88$ \\
12000 & $-1.30\times10^{0}$ & $6.56\times10^{-4}$ & $-5.07\times10^{-9}$ & - & - & - & $0.71$ \\
13000 & $-1.01\times10^{2}$ & $3.22\times10^{-2}$ & $-3.74\times10^{-6}$ & $2.08\times10^{-10}$ & $-5.44\times10^{-15}$ & $5.37\times10^{-20}$ & $1.15$ \\
14000 & $-2.37\times10^{2}$ & $6.86\times10^{-2}$ & $-7.35\times10^{-6}$ & $3.72\times10^{-10}$ & $-8.85\times10^{-15}$ & $7.88\times10^{-20}$ & $0.75$ \\
15000 & $2.07\times10^{1}$ & $-1.35\times10^{-3}$ & $3.34\times10^{-8}$ & - & - & - & $0.88$ \\
16000 & $-1.37\times10^{1}$ & $3.21\times10^{-3}$ & $-1.30\times10^{-7}$ & $1.60\times10^{-12}$ & - & - & $1.07$ \\
17000 & $1.32\times10^{1}$ & $-2.89\times10^{-3}$ & $2.96\times10^{-7}$ & $-1.05\times10^{-11}$ & $1.23\times10^{-16}$ & - & $1.03$ \\
\enddata
\end{deluxetable}

\end{document}